\documentstyle[graphics,IEEEmagn,babel]{article}

\newcommand{\bm}[1]{\mbox{\boldmath$#1$}}

\begin{document}
\oddsidemargin-0.6cm
\topmargin-2cm
\pagestyle{plain}
\footskip2cm

\baselineskip 10.875 pt

%Define symbol for fractions:
\def\tfrac#1#2{{\textstyle{#1\over#2}}}

\title{Dynamic electromagnetic response of three-dimensional Josephson junction arrays}

\author{J. Oppenl{\"a}nder, Ch. H{\"a}ussler  and N. Schopohl \\ 
{\small Institut f{\"u}r Theoretische Physik, Universit{\"a}t T{\"u}bingen, Auf der Morgenstelle 14, 
D-72076 T{\"u}bingen, Germany}}

\thanks{%
\vspace{-0.7cm}
\parindent0cm
\begin{tabular}{ll}
	E-mail addresses:	&joerg.oppenlaender@uni-tuebingen.de\\
				&christoph.haeussler@uni-tuebingen.de
\end{tabular}
}

\maketitle

\abstract{We present a theoretical study on the dynamical properties of three-dimensional
arrays of Josephson junctions. Our results indicate that such superconducting
networks represent highly sensitive 3D-SQUIDs having some major advantages in
comparison with conventional planar SQUIDs. The voltage response function of 3D-SQUIDs is 
directly related to the vector-character of external electromagnetic fields.
The theory developed here allows the three-dimensional reconstruction of a 
detected external field including phase information about the field variables. 
Applications include the design of novel magnetometers, gradiometers and particle detectors. }

\section{Introduction\label{sec1}}

Oscillators based on arrays of Josephson junctions operating in the voltage
mode have shown to be promising radiation sources in the mm- and sub-mm wavelength
range. One of the most remarkable features of such systems is that the microwave
power output increases with the number of array junctions while at the same
time the linewidth of the generated high-frequency voltage oscillation decreases
with this number [6]. The latter phenomena is not only interesting for
the application of the arrays as microwave sources but also for their application
as highly sensitive detectors. From this point of view, two- and three-dimensional
arrays of Josephson junctions represent multi-junction interferometers which
can be used e.g., as magnetometers, gradiometers or particle detectors. 

The capabilities of 2D- and 3D-Josephson arrays can be used successfully
if all array junctions operate in a coherent array mode, i.e., if they oscillate
strongly phase-locked [3,6]. The problem of phase-locking in 2D-arrays
is presently subject of intensive research and great experimental and theoretical
progress has been achieved using present-day integrated-circuit technology [1-3].
Since this technology is essentially two-dimensional there are up to now no
corresponding experimental and theoretical studies on 3D-arrays. However, 3D-arrays
of Josephson junctions may be promising candidates for highly sensitive detectors
which allow as a new quality the three-dimensional reconstruction of the detected
electromagnetic field including phase information about the field variables.
In view of recent developments in molecular beam epitaxy the fabrication
of 3D-arrays with appropriate junction parameters and array homogeneity seems
to be possible in the nearest future. 

\begin{figure}
{\centering \resizebox*{0.55\columnwidth}{!}{\includegraphics{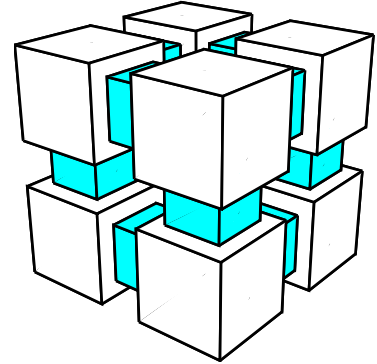}} \par}

\caption{\label{fig1}Schematic drawing of a 3D-network of Josephson junctions. The
network is built by superconducting islands (large cubes) connected by Josephson
junctions (smaller dashed cubes). }
\end{figure}

In this paper we present theoretical results on the dynamical properties of
a 3D-system of coupled Josephson junctions. A schematic drawing of the 3D-network
is shown in Fig.\ref{fig1}. It consists of superconducting islands (large cubes)
connected by Josephson junctions (smaller dashed cubes). This system is the
simplest three-dimensional configuration that guarantees coherent operation,
i.e., single frequency operation, if it is DC biased parallel to one of the
networks' axes (c.f. Fig.\ref{fig2}). If in the voltage state of the array
a magnetic field is applied to the network, the six intrinsically coupled four-junction
DC SQUIDs lying on the six faces of the cube are generating field dependent 
macroscopic current and voltage distributions which can be easily measured. 
From the AC Josephson effect and the flux quantization
condition the current and voltage distributions depend directly but nonlinearly
on the strength and the orientation of the external field.

\section{Network equations}

\begin{figure}
{\centering \resizebox*{0.85\columnwidth}{!}{\includegraphics{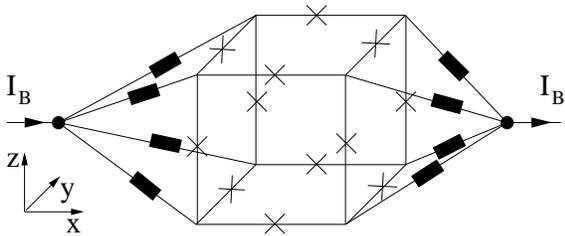}} \par}

\caption{\label{fig2}Equivalent circuit of the 3D-network. The Josephson junctions
(crosses) are connected by superconducting wires and the bias current \protect\( I_{B}\protect \)
is fed into and extracted from the system through ohmic resistors (black rectangles).}
\end{figure}

To get quantitative results about the networks' response function we will consider
a network built by junctions which can be described by the RCSJ-model, e.g.,
externally shunted \( Nb-Al0_{x}-Nb \) junctions. However, if other SIS-type junctions
are used the results remain qualitatively valid. Fig.\ref{fig2} shows the equivalent
circuit for the 3D-network which is DC biased along the x-direction. It consists
of Josephson junctions (crosses) which are connected by superconducting wires.
The bias current \( I_{B} \) is fed into and extracted from the network through
ohmic resistors (which must not necessarily be equal). This method of biasing
preserves well defined boundary conditions for the networks dynamics and removes
some of the ambiguities due to the symmetry of the system. Since the bias current
is flowing parallel to the x-axis there are two groups of junctions with different
dynamical behavior. The junctions lying parallel to the bias current (active
junctions) switch into their voltage state for \( I_{B}>I_{c,A}(\bm {f}_{\mbox {ext}}) \),
where \( I_{c,A}(\bm {f}_{\mbox {ext}}) \) is the critical current of the array
depending on the externally applied flux \( \bm {f}_{\mbox {ext}}=(f_{x},\, f_{y},\, f_{z}) \).
The junctions lying perpendicular to the bias current (passive junctions) do
in general not switch into the voltage state but show librations (semirotations)
of their Josephson-phase differences. The amplitude of these librations is directly
related to the strength and orientation of the external field. If the values of the ohmic resistors
at the current in- and output do not differ too much, the passive junctions are in the 
zero-voltage state and the only frequency present in the system is the driving frequency 
defined by the active network junctions. This prevents chaotic or aperiodic dynamics of the 
junction phases.

In the temporal gauge the variables characterizing the dynamics of the junction
network are the gauge invariant phase differences [4,5]. If we denote
the vector of the 12 network variables by \( \bm {\chi }(t) \) and use the
usual reduced units [1] the system of coupled network equations can be
written symbolically as
\begin{eqnarray}
\beta \, \bm {\ddot{\chi }}+\bm {\dot{\chi }}+\sin \, \bm {\chi } & = & \lambda \, \left[ L^{-1}\, \bm {\chi }+F\, \bm {f}_{\mbox {ext}}\right] \label{1} \\ 
&  & +E \, \bm {j}_{\mbox {rf}} \nonumber \\
 &  & +G(\bm {r_{B}})\, I_{B}\nonumber \\
 &  & +D(\bm {r_{B}})\, \bm {\dot{\chi}},\nonumber 
\end{eqnarray}
 where \( \lambda =\Phi _{0}/(2\pi \, \mu _{0}\, i_{c}\, a) \) is the magnetic
penetration depth, \( L \) the inductance matrix of the array, \( F \) the
matrix of the distribution of the external flux $\bm {f}_{\mbox {ext}}$ over the different network meshes
and  \( E \) the matrix which describes the distribution of the currents $\bm {j}_{\mbox {rf}}$ induced
by the oscillating electric field vector of an incoming external wave. The elements of \( E \) and \( F \) depend 
on the orientation of the external electromagnetic field. 
\( G \) and \( D \) are the matrices which define the boundary conditions
depending on the values of the eight input and output resistors \( \bm {r_{B}}=\left\{ r_{B,i}\right\} . \)
The single junction parameters are the McCumber parameter \( \beta  \), the
critical current \( i_{c} \) and the normal resistance \( r \). \( \Phi _{0} \)
is the magnetic flux quantum and \( a \) the lattice spacing measuring the
distance between the centers of adjacent superconducting islands. The inductance
matrix \( L \) includes all inductances present in the array and
depends on the geometry of the network and the inductances of the single junctions.
We determine the coefficients of \( L \) by assuming an array geometry similar
to that shown in Fig.\ref{fig1} and by using junction inductances which are
typical for externally shunted junctions. Changing the geometry of the network
or the value of the junction inductances, however, influences our results only
slightly.

The network equation (\ref{1}) clearly shows the dependence of the time evolution
of the network phases \( \bm {\chi }(t) \) on the applied quasi-static external magnetic
field which can approximately be expressed as 
\(\bm {f}_{\mbox {ext}}=\bm {B_{\mbox {ext}}}(x,\, y,\, z)\, a^{2}/\Phi _{0} \),
where \( \bm {B_{\mbox {ext}}} \) is the magnetic field vector of the external
field. The sensitivity of the network depends on the lattice spacing \( a \)
and reaches its maximum for vanishing magnetic coupling within the network,
i.e., for infinite magnetic penetration depth \( \lambda \rightarrow \infty . \)
For \( \lambda \rightarrow 0 \), however, the network response vanishes totally.
The sensitivity of the network dynamics with respect to distinct directions
of the magnetic field, i.e., the anisotropy of the response function, can be
manipulated by an appropriate choice of the input and output resistors. 

The device possesses in principle two modes of operation. The DC mode with \( I_{B}<I_{c,A} \)
and the AC mode with \( I_{B}>I_{c,A} \). For subcritical bias currents \( I_{B}<I_{c,A} \),
a constant external field induces constant loop currents in the meshes of the network
and the resulting current distribution within the array is a superposition of
these loop currents with the bias current. The critical array current \( I_{c,A}(\bm {f}_{\mbox {ext}}) \)
is a function of the strength of the external field and the orientation of this
field. However, because of the many degrees of freedom of the system the response
function is not unique. There exists in general for any given \( \bm {B_{\mbox {ext}}} \)
a whole set of different equilibrium current distributions in the network on
which the critical array current depends [5,7]. By a detailed theoretical
analysis it is possible to formulate selection rules which describe the external
field depending transitions between the different distributions. Although there
occur in the region of subcritical bias currents very interesting
dynamical phenomena, the scenario becomes very complicated and will be presented
elsewhere [7]. 

In the following we will restrict our treatment to the AC mode of the network.
For \( I_{B}>4\, i_{c} \) the active array junctions show persistent voltage
oscillations whose frequency is directly proportional to the voltage drop. If
the constant voltage drop originating from the input and output resistors
is subtracted, the averaged voltage drop across the array is for \( I_{B}>I_{c,A} \)
given by \( V_{0}=1/4<\sum _{x}\, \dot{\chi }_{x}(t)> \), where the summation
runs over the four active network junctions lying parallel to the x-axis and
\( <\ldots > \) denotes time averaging over one oscillation period. Due to
flux quantization, the network currents induced by the external field are converted
by the network dynamics into oscillating loop currents in each network mesh.
The voltage oscillation induced by these oscillating loop currents interferes
with the voltage oscillation of the active junctions, so that the averaged voltage
drop \( V \) across the whole array becomes a function of the strength and
the orientation of the external field. 

\begin{figure}
{\centering \resizebox*{1.05\columnwidth}{!}{\includegraphics{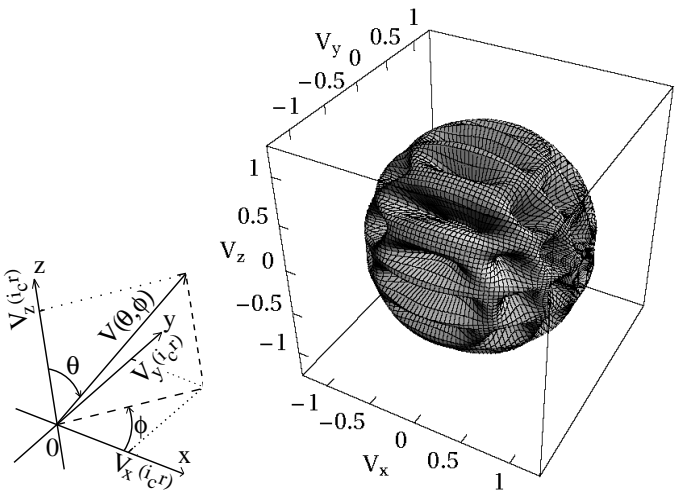}} \par}
\caption{\label{fig3}Spherical representation of the macroscopic voltage response function
of the 3D-network. The time averaged voltage drop \protect\( V_{0}\protect \)
across the network is plotted in the form  \( \bm {V}(\theta ,\phi )=V_{0}\, \bm {\hat{e}}_{\theta ,\phi}\)
as a function of the direction \(\bm {\hat{e}}_{\theta ,\phi}\) of an applied constant
external magnetic field \protect\( \bm{B_{\mbox {ext}}}=B_{0,\mbox {ext}}\, \bm{\hat{e}}_{\theta ,\phi} \protect\) 
inducing a flux  \( \left| \bm {f}_{\mbox {ext}}\right|  \)\( =B_{0,\mbox {ext}}\, a^{2}/\Phi _{0}=3\). 
The system of coordinates on the l.h.s shows the definition of the spherical angles \protect\( \theta \protect \) 
and \protect\( \phi .\protect \) The
origin lies in the center of the cube such that \protect\( V_{0}=\left| \bm {V}(\theta ,\, \phi )\right| \protect \).}

\end{figure}

\section{Results}

By numerically integrating the network equation (\ref{1}) we computed the macroscopic
voltage response function of the 3D-network for various different sets of junction
and array parameters. In the following we present typical results. Fig.\ref{fig3}
shows a spherical plot of the voltage response function for \( \beta =0.5, \)
\( \lambda =5 \), \( I_{B}=4.4\, i_{c} \) and a homogeneous magnetic field
\( \bm {B_{\mbox {ext}}}(x,\, y,\, z)=B_{0,\mbox {ext}}\, \bm {\hat{e}}_{\theta ,\phi } \)
which induces a flux \( \left| \bm {f}_{\mbox {ext}}\right|  \)\( =B_{0,\mbox {ext}}\, a^{2}/\Phi _{0}=3. \)
Here \( \bm {\hat{e}}_{\theta ,\phi } \) is a vector with unit length pointing
in the direction given by the spherical coordinates \( \theta  \) and \( \phi  \)
and the origin is put in the center of the 3D-array (c.f. Fig.\ref{fig3}).
The voltage response function is plotted in the form \( \bm {V}(\theta ,\phi )=V_{0}\, \bm {\hat{e}}_{\theta ,\phi }=(V_{x},V_{y},V_{z}) \),
where \( V_{0} \) is the macroscopic averaged voltage drop across the array
measured in units of \( i_{c}\, r \). The eight input and output resistors
(c.f. Fig.\ref{fig2}) all have the same value so that, corresponding to the
symmetry of the network, the voltage response function shows a periodicity with
period \( \pi /2 \) along each closed intersection curve with the x-z-, x-y-
and y-x-plane, respectively, and a reflection symmetry with respect to each
integer multiple of \( \pi /4. \)

\begin{figure}[h]
{\centering \begin{tabular}{c}
(a)\\[0.3cm]
\resizebox*{0.95\columnwidth}{!}{\includegraphics{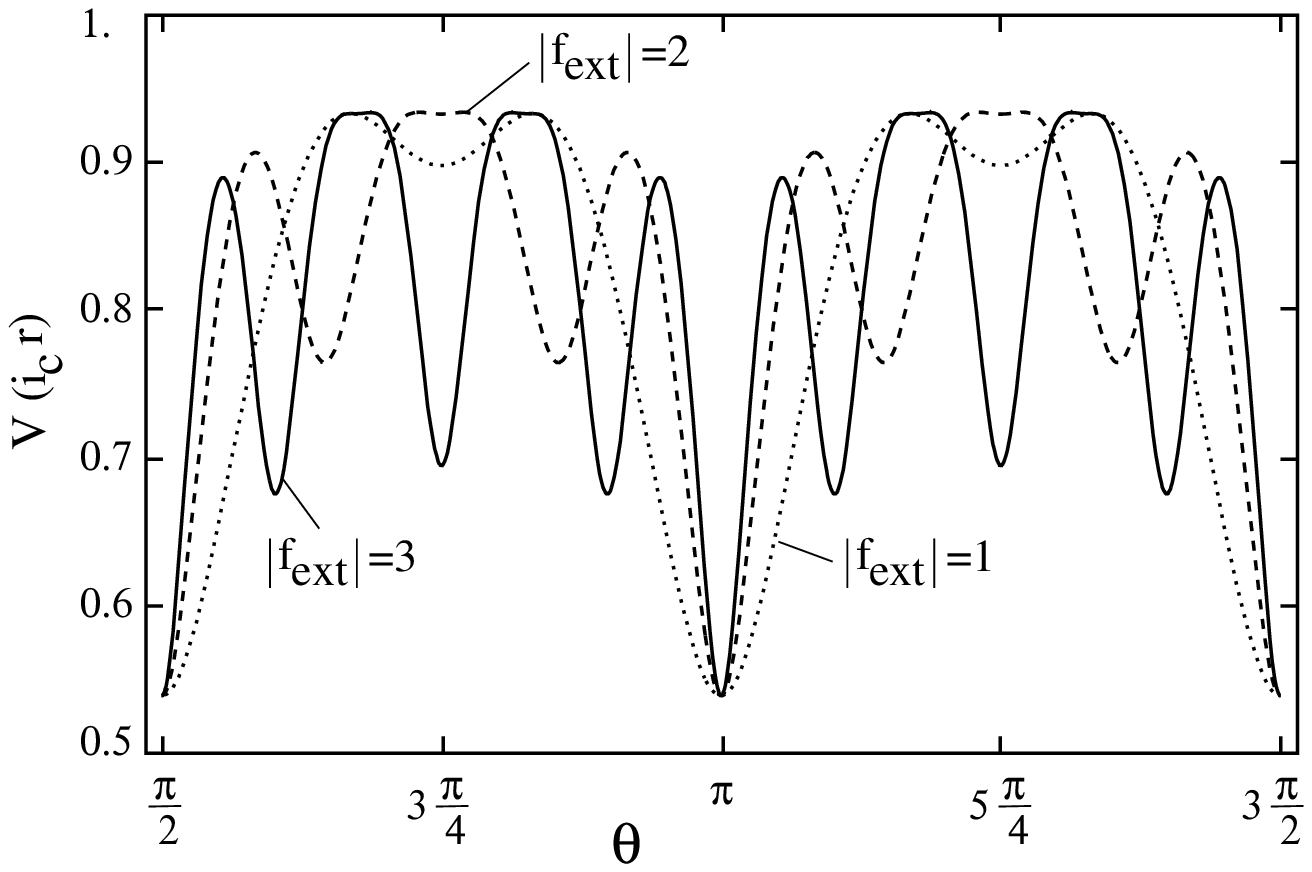}}\\[0.3cm]
(b)\\[0.3cm]
\resizebox*{0.96\columnwidth}{!}{\includegraphics{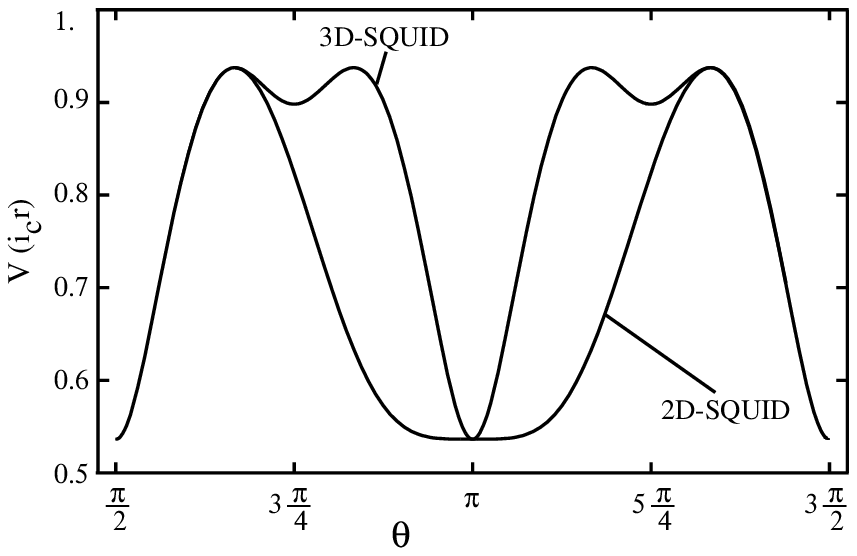}} \\
\end{tabular}\par}
\caption{\label{fig4} (a) Voltage response function \protect\( \bm {V}(\theta ,\phi =\pi /2)\protect \)
in the y-z-plane (\protect\( x=0\protect \)) for \protect\( \theta =\left[ \pi /2,\, 3\pi /2\right] \protect \)
and different values of the external applied magnetic field \protect\( \bm {B_{\mbox {ext}}}\protect \)
which induces a flux of \protect\( \left| \bm {f}_{\mbox {ext}}\right| =1\protect \)
(dotted line), \protect\( \left| \bm {f}_{\mbox {ext}}\right| =2\protect \)
(dashed line) and \protect\( \left| \bm {f}_{\mbox {ext}}\right| =3\protect \)
(solid line). (b) Comparison of the voltage response function of the 3D-SQUID (c.f. Fig.\ref{fig2}) and a 
conventional 2D-SQUID which consists of a superconducting loop that contains two junctions. The bias current
per active junction is for both configurations $1.1 \, i_c$. }
\end{figure}

In Fig.\ref{fig4}(a) intersection curves in the z-y-plane, i.e., \( \phi =\pi /2 \),
are plotted for \( \theta =\left[ \pi /2,\, 3\pi /2\right]  \) and different
values of \( \left| \bm {f}_{\mbox {ext}}\right|  \). The symmetry of the voltage
response functions with respect to integer multiples of \( \pi /4 \) is clearly
observable and the sensitivity of the network reaches its maximum around integer
multiples of \( \pi /2. \) For such values of \( \theta  \) the field vector
of the external magnetic field lies directly perpendicular to one of the faces
of the cube and the induced currents in general become maximal in the direction parallel
(and antiparallel) to the bias current. In this case the externally induced
voltage drop becomes also maximal. The number of local maxima and minima of
the voltage response function is strongly related to the applied flux \( \left| \bm {f}_{\mbox {ext}}\right|  \)
and can be determined numerically. 

The resolving power of 3D-SQUIDs with respect to the strength of external magnetic fields
is comparable to the resolving power of a conventional 2D-SQUID, i.e. a SQUID loop with two junctions.
The angular resolution of 3D-SQUIDs, however, is orders of magnitudes better than for 2D-SQUIDs.
For external magnetic fields inducing a flux \(|f_{ext}|\) in the order of one, 
the transfer factor \(|\partial V / \partial \theta|\) lies around \(0.02 \,i_c \, r / rad \) which is for typical
junctions in the range of some \(10\,\mu  \)V\(/rad\), such that, e.g., an angle of
\(10^{-2}\,rad\) should be easily resolvable. The resolving power with respect to the strength 
and the resolving power with respect to the direction of an external field can be further increased
by using networks which consist of several cubes. In this case the resolution powers increase proportional
to the number of cubes. 

Fig.\ref{fig4}(b) shows for comparison the angular resolution of the 3D-junction array and a
corresponding conventional 2D-SQUID. Both configurations
are considered to have identical parameters. If the applied magnetic field lies perpendicular 
to the 2D-SQUID plane, i.e. $\theta=\pi$, the device is very insensitive to variations of $\theta$.
This insensitivity, however, has advantages if only the strength of the magnetic field is to be measured.
In contrast to this, the 3D-array's response is very sensitive to variations of the direction of the 
external magnetic field. Therefore, the 3D-network has the advantage that simultaneously the strength 
and the direction of the magnetic field can be measured very precisely. 

Three-axis SQUIDs, which consist of three independent operating 2D-SQUIDs are not directly comparable with
3D-SQUIDs. For three-axis SQUIDs the three independent 
voltage response functions are computationally postprocessed and the information about strength and 
orientation of the magnetic field vector is computationally extracted. 
In contrast to this, the single voltage response function of 3D-SQUIDs provides all information
at once and no further postprocessing is needed.

\begin{figure}[h]
{\centering \begin{tabular}{cc}
(a)&
(b)\\
\resizebox*{0.45\columnwidth}{!}{\includegraphics{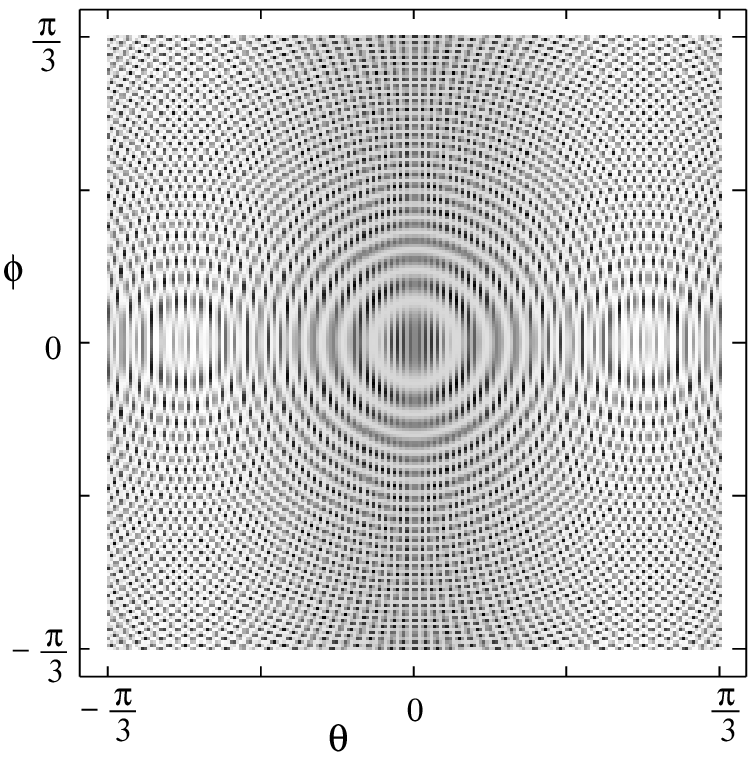}} &
\resizebox*{0.46\columnwidth}{!}{\includegraphics{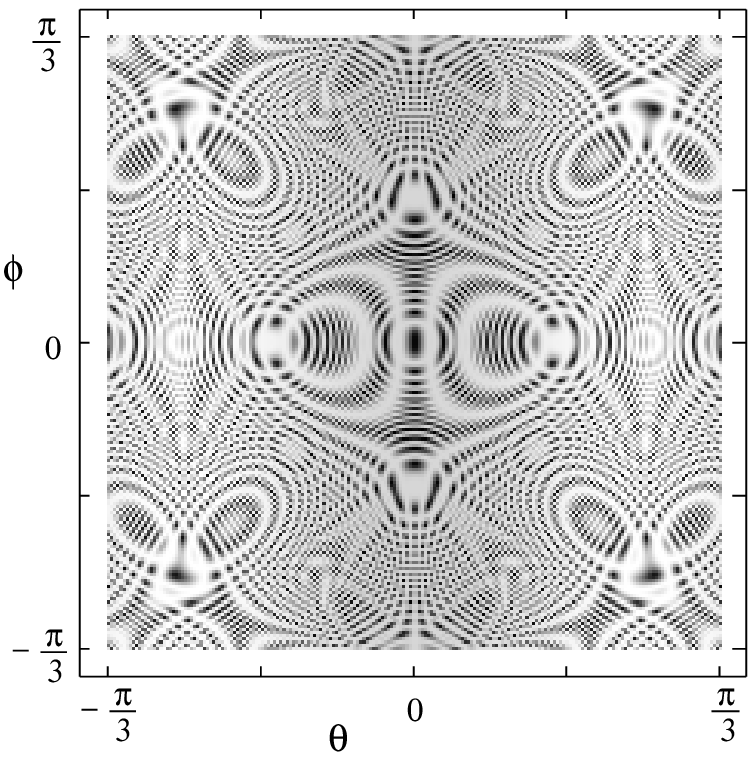}} \\
\end{tabular}\par}
\caption{\label{fig5}Density plot of the voltage response function for a solid angle
with \protect\( \theta =\left[ -\pi /3,\, \pi /3\right] \protect \) and \protect\( \phi =\left[ -\pi /3,\, \pi /3\right] \protect \).
Black dots indicate minima of the response function and white areas maxima (c.f.
Fig.\ref{fig4}). The external magnetic field induces in (a) a flux of \protect\( \left| \bm {f}_{\mbox {ext}}\right| =50\protect \)
and in (b) \protect\( \left| \bm {f}_{\mbox {ext}}\right| =200\protect \).}
\end{figure}

For increasing \( \left| \bm {f}_{\mbox {ext}}\right|  \) the number of local
maxima and minima of the voltage response function grows. Fig.\ref{fig5} shows
density plots of the response function for a solid angle \( \Delta \Omega =\Delta \theta \, \Delta \phi  \)
with \( \Delta \theta =\Delta \phi =2\pi /3 \), and (a) \( \left| \bm {f}_{\mbox {ext}}\right| =50 \)
and (b) \( \left| \bm {f}_{\mbox {ext}}\right| =200 \). The network para\-meters
are the same as for Figs.(\ref{fig3}) and (\ref{fig4}). The minima (black
dots) and maxima (white areas) form significant patterns which are quite similar
to optical interference patterns. However, according to Eq.(\ref{1}) these
patterns are a result of nonlinear interactions and a trace of their nonlinear
origin are the complex local structures occurring in Fig.\ref{fig5}(b). By
evaluating the voltage response patterns, the strength of the external magnetic
field and, up to the symmetry redundancy, the orientation of this field can
be determined very precisely. The symmetry redundancy can be removed by an appropriate
choice of the input and output resistors. In this case, however, the resistors
possess slightly different values and the voltage response of the network becomes
much more complicated and more difficult to interpret [7]. 

\begin{figure}[ht]
{\centering \begin{tabular}{c}
(a)\\
\resizebox*{0.60\columnwidth}{!}{\includegraphics{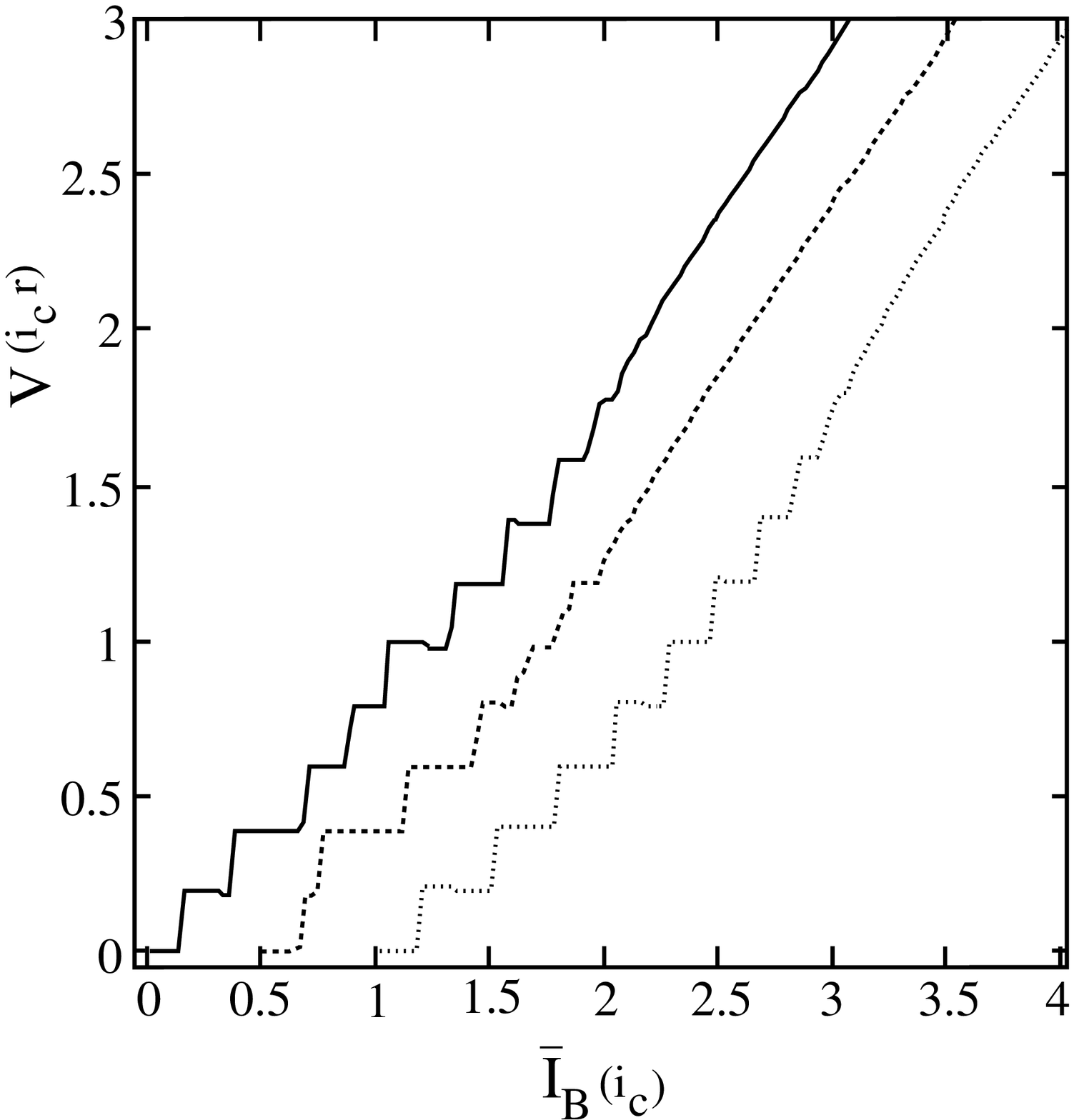}}\\
(b)\\
\resizebox*{0.60\columnwidth}{!}{\includegraphics{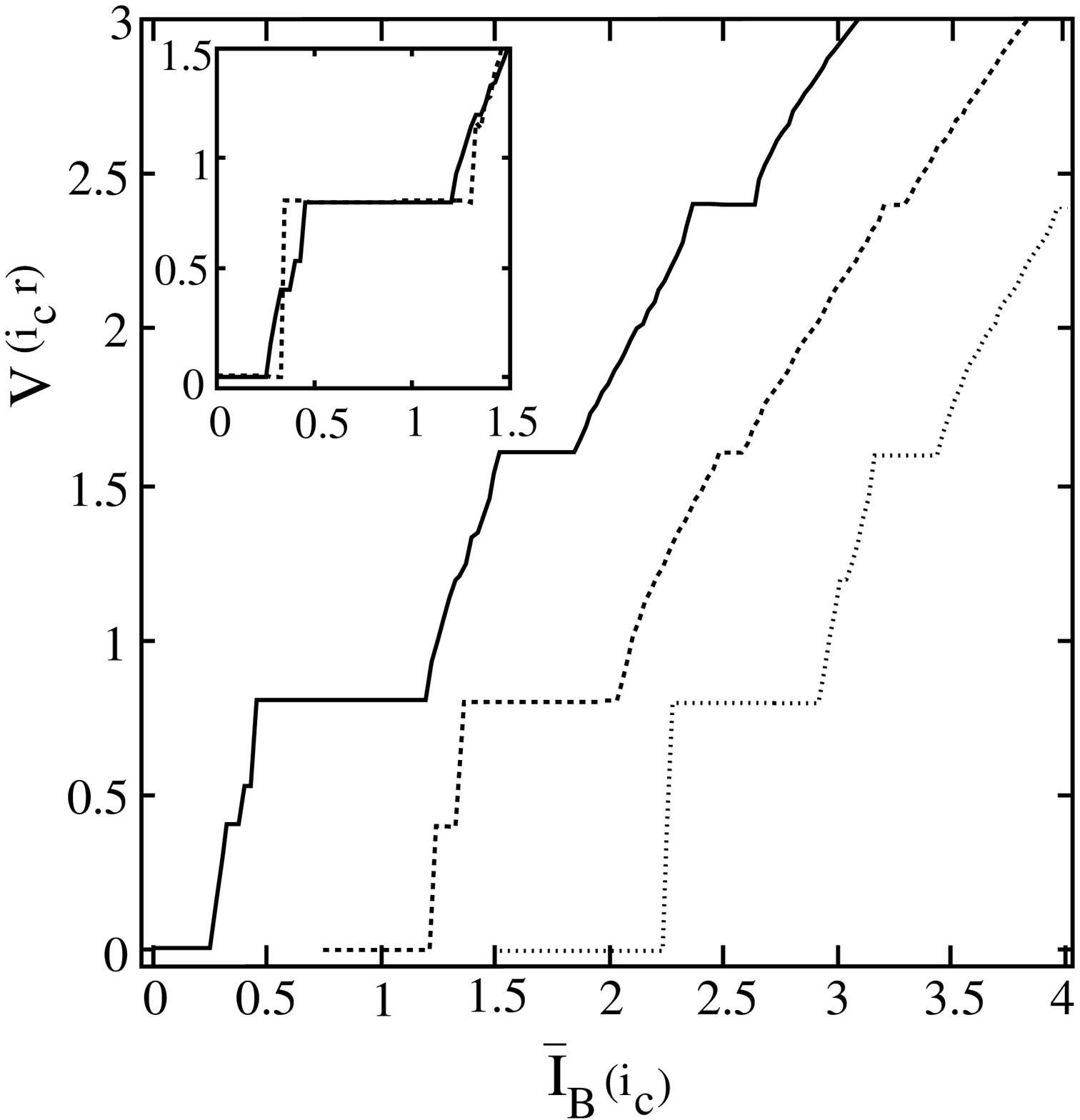}} \\
\end{tabular}\par}
\caption{\label{fig6}Current-voltage characteristics of the 3D-network for a time dependent
external electromagnetic field. The external microwave is linearly polarized parallel to the bias current
and the direction of the magnetic field vector is in (a) and (b) \protect\( (\theta ,\, \phi )=(0,\, \pi /2)\protect \)
(solid curve) and \protect\( (\theta ,\, \phi )=(\pi /4,\, \pi /2)\protect \)
(dashed curve).  The frequencies of the external microwave are
\protect\( \nu =0.2\, i_{c}\, r\protect \) for (a) and \protect\( \nu =0.8\, i_{c}\, r\protect \)
for (b). The amplitudes of the externally induced flux and of the induced current
are for both figures \protect\( \left| \bm {f}_{\mbox {ext}}\right| =\left| \bm {j}_{\mbox {rf}}\right| =1\protect \).
 For comparison also the I-V-curves for negligible influence
of the magnetic field (dotted curves) are plotted. The dashed and dotted curves in (a) have been 
offset by 0.5 units along the x-axis and in (b) by 0.75 units. The inset of (b) shows the
splitting of the characteristics for a linearly polarized microwave \( (\theta ,\, \phi )=(0,\, \pi /2)\protect \) 
(solid curve) and a microwave that is circularly polarized in the z-x-plane (dashed curve).}
\end{figure}

If the parameters of the array junctions and the dimension of the 3D-network
are chosen appropriately, the systems are also able to detect time dependent
electromagnetic fields with wavelengths which lie typically in the mm-range.
For time dependent external fields the induced flux becomes time dependent \( \bm {f}_{\mbox {ext}}=\bm {f}_{\mbox {ext}}(t) \)
and the currents \( \bm {j}_{\mbox {rf}}(t) \) are induced by the oscillating electric
field vector of the external wave (c.f. Eq.1). In contrast to 2D-configurations, for which the influence
of the oscillating magnetic field is in general negligible, for the 3D-network
both field vectors contribute to the response function of the array if the lattice
spacing \( a \) is chosen appropriately. Fig.\ref{fig6} shows current-voltage
characteristics of the 3D-network for different fixed directions of the external
microwave and array parameters \( \beta =0.5 \) and \( \lambda =5 \). The
microwave is assumed to be linearly polarized parallel to the bias current and
the amplitudes of the externally induced flux and of the induced current \( \bm {j}_{\mbox {rf}} \)
are \( \left| \bm {f}_{\mbox {ext}}\right| =\left| \bm {j}_{\mbox {rf}}\right| =1 \).
The frequency of the external microwave is \( \nu =0.2\, i_{c}\, r \) in Fig.\ref{fig6}(a)
and \( \nu =0.8\, i_{c}\, r \) in Fig.\ref{fig6}(b), and \( \bar{I}_{B}=I_{B}/4 \)
is the normalized bias current. In each of both figures the I-V-curves for the magnetic
field directions \( (\theta ,\, \phi )=(0,\, \pi /2) \)
(solid curve) and \( (\theta ,\, \phi )=(\pi /4,\, \pi /2) \) (dashed curve)
are plotted, and, for comparison, the I-V-curve for negligible influence of
the magnetic field (dotted curve).

It is clearly observable that the I-V-characteristics
depends on the direction of the incoming wave and that the magnetic part is
indeed relevant. In general, the distribution of the Shapiro steps and their width
differ significantly from those corresponding to a single junction. In addition, it can be observed that for some 
characteristics the step plateaus show an unusual fine structure. The first, fifth and seventh step on the left characteristics 
in  Fig.\ref{fig6}(a) show such structures. These fine structures are according to Eq.(1) 
implied by the different contributions of the electric and the magnetic field to the network dynamics.
We observed that on the Shapiro steps the network junctions first lock to the electric part of the external microwave 
and then for increasing bias current eventually also to the magnetic part. This can change the current 
distributions within the network and implies a slight decrease in the averaged voltage drop and therefore 
in the network frequency.

The 3D-network dynamics governed by Eq.(1) includes all informations about the field variables of 
the external microwave. By a detailed analysis it is therefore possible to extract from the 
macroscopic current and voltage response functions of the network the information about the phase 
relationship within the incoming wave. 

By assuming the simple model of a circularly polarized incoming microwave
we can show that the voltage response function of the 3D-network is directly affected
by the helicity of the external field. The inset of Fig.\ref{fig6}(b) shows
the I-V-characteristics near the critical array current \( I_{c,A} \) for \( (\theta ,\, \phi )=(0,\, \pi /2) \)
for a microwave that is linearly polarized in the x-direction (solid curve)
and a microwave that is circularly polarized in the z-x-plane (dashed curve).
The splitting between the critical array current \( I_{c,A} \) of the two modes
lies in the range of several \( 0.01\, i_{c}\, r \) and the shape of the characteristics
differ significantly. This result already indicates that the 3D-network can
operate phase-sensitively and that the modulations in the voltage response function
caused by differently polarized external microwaves should be easily observable
in experiments.

\section{Conclusions}

By presenting various theoretical results we have shown that 3D-networks of
Josephson junctions represent ultrasensitive 3D-SQUIDs which can be useful for
a number of different applications like magnetic field sensors, video detectors
or mixers. Especially the capability of the networks to operate direction-sensitively
and to detect incoming microwaves phase-sensitively is a novel quality and may 
allow the design of a new generation of superconducting devices.

\section*{Acknowledgment}
Financial support by the Forschungsschwerpunktprogramm des Landes Baden-W\"urttemberg is gratefully acknowledged.

\end{document}